\documentclass[universe,article,accept,pdftex,moreauthors]{Definitions/mdpi} 

\firstpage{1} 
\pubvolume{1}
\issuenum{1}
\articlenumber{0}
\pubyear{2025}
\copyrightyear{2025}
\externaleditor{Firstname Lastname}
\datereceived{10 December 2024} 
\daterevised{25 January 2025} % Comment out if no revised date
\dateaccepted{26 January 2025} 
\datepublished{ } 
\hreflink{https://doi.org/} % If needed use \linebreak
\def\prd{Phys.~Rev.~D}       % Physical Review D
\def\actaa{Acta Astron.}      % Acta Astronomica
\def\apj{Astrophys. J.}
\def\pasj{\ref@jnl{PASJ}}               % Publications of the ASJ

\newcommand{\gttup}{g^{tt}}
\newcommand{\gtpup}{g^{t\phi}}
\newcommand{\gppup}{g^{\phi \phi}}

\newcommand{\gtp}{g_{t\phi}}
\newcommand{\gpp}{g_{\phi \phi}}
\newcommand{\utup}{U^t}
\newcommand{\upup}{U^\phi}
\newcommand{\ut}{U_t}

\newcommand{\at}{A_t}
\newcommand{\aphi}{A_\phi}

\newcommand{\Tab}{T^{\alpha \beta}}
\newcommand{\Tabmat}{T^{\alpha \beta}_{\text{MAT}}}
\newcommand{\Tabem}{T^{\alpha \beta}_{\text{EM}}}
\newcommand{\Fab}{F^{\alpha \beta}}
\newcommand{\Fabext}{F^{\alpha \beta}_{\text{EXT}}}
\newcommand{\Fabint}{F^{\alpha \beta}_{\text{INT}}}

\newcommand{\At}{A_{t}}
\newcommand{\Ap}{A_{\phi}}
\usepackage{siunitx}

\DeclareSIUnit[quantity-product = {}]
\gauss{\text{ G}}

\Title{Non-Keplerian Charged Accretion Disk Orbiting a Black Hole~Pulsar}

\TitleCitation{Non-Keplerian Charged Accretion Disk Orbiting a Black Hole Pulsar}

\Author{Audrey %MDPI: Please carefully check the accuracy of names and affiliations.
 Trova *%MDPI: Newly added information based on corresponding author's email. Please confirm.
\orcidA{} and Eva Hackmann \orcidB{}}

\AuthorNames{Audrey Trova and Eva Hackmann}

\AuthorCitation{Trova, A.; Hackmann, E.}
\address[1]{University of Bremen, Center of Applied Space Technology and Microgravity (ZARM), 28359 Bremen, Germany; %MDPI: Newly added information. Please confirm.
eva.hackmann@zarm.uni-bremen.de %MDPI: We added the email addresses here according to those submitted online at susy.mdpi.com. Please confirm.
}

\corres{\hangafter=1 \hangindent=1.05em \hspace{-0.82em}Correspondence: audrey.trova@zarm.uni-bremen.de}

\abstract{Recent studies have focused on how spinning black holes (BHs) within a binary system containing a strongly magnetized neutron star, then immersed in external magnetic fields, can acquire charge through mechanisms like the Wald process and how this charge could power pulsar-like electromagnetic radiation. Those objects called ``Black hole pulsar'' mimic the behaviour of a traditional pulsar, and they can generate electromagnetic fields, such as magnetic dipoles.
Charged particles within an accretion disk around the black hole would then be influenced not only by the gravitational forces but also by electromagnetic forces, leading to different geometries and dynamics. 
In this context, we focus here on the interplay of the magnetic dipole and the accretion disk. We construct the equilibrium structures of non-conducting charged perfect fluids orbiting Kerr black holes under the influence of a dipole magnetic field aligned with the rotation axis of the BH. The dynamics of the accretion disk in such a system are shaped by a complex interplay between the non-uniform, non-Keplerian angular momentum distribution, the black hole's induced magnetic dipole, and the fluid's charge. We show how these factors jointly influence key properties of the disk, such as its geometry, aspect ratio, size, and rest mass density.}


\begin{document}

%%%%%%%%%%%%%%%%%%%%%%%%%%%%%%%%%%%%%%%%%%
%\setcounter{section}{-1} %% Remove this when starting to work on the template.
% \section{How to Use this Template}

% The template details the sections that can be used in a manuscript. Note that the order and names of article sections may differ from the requirements of the journal (e.g., the positioning of the Materials and Methods section). Please check the instructions on the authors' page of the journal to verify the correct order and names. For any questions, please contact the editorial office of the journal or support@mdpi.com. For LaTeX-related questions please contact latex@mdpi.com.%\endnote{This is an endnote.} % To use endnotes, please un-comment \printendnotes below (before References). Only journal Laws uses \footnote.

% The order of the section titles is different for some journals. Please refer to the "Instructions for Authors” on the journal homepage.

\section{Introduction}
New and improved astrophysical observation techniques have, in recent years, collected overwhelming evidence for the existence of black holes, including, e.g., observations of supermassive black holes in active galactic nuclei \cite{PadovaniReview2017}, X-ray observations of stellar-mass black holes \cite{Miller_Jones_2021}, the observation of stars around Sgr A* \cite{Gillessen_2017,Reid_2020}, the first images of the shadows of black holes \cite{EHT2019_I,EHT_2022}, and the direct observation of gravitational waves \cite{Abbott_2016,Abbott_2021}. Although black holes in vacuum are ``clean'' objects that are characterized by mass, angular momentum, and charge, only due to the no-hair theorem, considerable additional effects in the vicinity of black holes arise due to their interaction with the environment. Traditionally, it has been argued that astrophysical black holes are uncharged, as any electric charge would quickly be neutralized due to the selective accretion of oppositely charged particles from the environment, or even from pair production \cite{Gibbons1975,Eardley1975}. However, as pointed out by Wald \cite{Wald74}, and recently revisited by Komissarov \cite{Komissarov2022}, a spinning black hole may acquire a stable net electric charge if immersed in an external asymptotically uniform magnetic field. \mbox{Levin et al.~\cite{Levin2018}} studied the possibility that such a magnetic field is provided by a neutron star companion in a binary system. Another possibility would be a more distant magnetar, maybe orbiting together with a stellar-mass black hole around Sgr A*, or even galactic-scale magnetic fields. The resulting electric charge of the black hole itself is proportional to the magnetic field strength \cite{Wald74} and likely to converge to a small value over time. However, the electric charge distribution emerges within the surrounding environment; the charges can separate and play a significant role in the interaction with the global magnetic field. A spinning electric charge naturally creates a magnetic dipole. As argued by Levin et al. \cite{Levin2018}, such a black hole actually resembles a pulsar, with spin, a magnetic field, and maybe even a strong enough electric field to create a magnetosphere. Another possibility is that a magnetic dipole field may be created by the accretion disk itself \cite{Prasanna1978,Kovar2016,Schroven2018}.

In this paper, we discuss the possibility of a long-lived black hole pulsar---or generally, a black hole with a dipole electromagnetic field---surrounded by an accretion disk. Such a gravitational and electromagnetic background generally plays an important role in the structure of accretion flows \cite{Pariev_2003,Sadowski_2016}. As the matter in accretion disks is generally charged, here, we are interested in the impact of such a dipole field on the overall general features of the disk. To understand the basic properties without focusing too much on the variability, as a starting point, we consider an equilibrium configuration that mimics the averaged dynamics of the disk. Therefore, we use here an analytical approach that combines charged accretion disks as constructed in \cite{Kovar2016,Schroven2018} as a generalization of the classical Polish doughnut model (see, e.g.,~\cite{Kozlowski1978,AbramowiczLRR}) with a non-constant angular momentum profile similar to that in ref.~\cite{Qian2009}. We aim to extract the physical characteristics of the accretion disk and to study the interplay of the different involved parameters of the black hole spin, the electromagnetic interaction, and the angular momentum profile in the disk. Our results can then serve as a basis to compute spectra, e.g., to model black hole images similar to \cite{Vincent2015} or serve as initial conditions in resistive GRMHD simulations.  

To facilitate an analytical approach, we need to make some simplifying assumptions. Firstly, we assume that the whole setup is stationary and axially symmetric and the spinning black hole is described by the Kerr metric. 
Moreover, we assume a physically reasonably hierarchy in the involved electromagnetic fields: the dipole magnetic field $B$ associated with the black hole and the electromagnetic field associated with the charged accretion disk. As discussed previously, the black hole can sustain a magnetic dipole field through its rotation. The magnitude of such system can exceed $\qty{e8}{\gauss}$ above the pulsar limit and reach strengths similar to those of neutron stars ($\qty{e12}{\gauss}$). However, this is still quite weak if considering the impact on the spacetime geometry. As, e.g.,~discussed in  \cite{2010PhRvD..82h4034F}, the magnetic field has a noticeable influence on the curvature of spacetime if $GB^2/c^4 \sim r_g^{-2}$, where $G$ is the gravitational constant, $c$ is the speed of light, and $r_g$ is the gravitational radius. Therefore, to be considered a test field, the strength of the magnetic field should not exceed 
    \begin{equation}
        B \sim 10^{19}\, \frac{M_{\odot}}{M}\, \text{G},
    \end{equation}
where $M_{\odot}$ is the mass of the Sun, and $M$, the mass of the BH. The corresponding electric charge of the spinning black hole, acquired through the process described by Wald \cite{Wald74}, can then be estimated to not exceed \cite{Wald74,Komissarov2022}
\begin{align}
    Q < 2Ba \leq 2BM \sim 0.17 M\,. 
\end{align}

    Here, $Q$ is the electric charge parameter of the BH\endnote{Note that $Q$ only appears squared in the Kerr--Newman metric, in contrast to $a$. In SI units, $Q = \sqrt{\frac{G}{4\pi \epsilon_0 c^4}} Q_{\rm SI} \sim 8.6 \times 10^{-18} Q_{\rm SI}$. For Sgr A*, an observational bound is $Q_{\rm SI} \leq 3 \times 10^{8} \, \rm C$ \cite{Zajacek2019}.}, and $a$, its spin parameter, assumed to satisfy the black hole bound $a\leq M$. As the magnetic field has a strength of about $\qty{e8}{\gauss}$ to $\qty{e12}{\gauss}$, we can therefore safely neglect the influence of both the magnetic field and the corresponding electric field on the spacetime geometry. Still, the magnetic dipole field can be assumed to be much stronger than the internal fields produced by the circling charged fluid disk, which we may therefore be neglected as well in our treatment.    

Non-Keplerian disk models have been extensively investigated in recent decades. Among them, the thick disk model \cite{1976ApJ...207..962F,AbrJaSi78} and its magnetized version \cite{Komissarov06} have been studied in various spacetime geometries. This paper, which is a charged version of the thick disk model, is built upon several previous studies \cite{KovarTr14,Kovar2016,2020Trova,2024IJMPD..3350112T}, among others, with foundational work tracing back to \cite{Kovar11}. In particular, the work presented here is an extension of \cite{Schroven2018}, where the specific angular momentum was assumed to be constant. These studies have investigated the equilibrium of charged accretion disks under various combinations of spacetime geometries due to the (un)charged central mass (e.g., Kerr, Schwarzschild), external magnetic fields (e.g., uniform, dipole), and accretion disk rotation laws (e.g., rigid rotation, constant angular momentum). They have demonstrated that equatorial solutions comprising single or double toroidal structures can be constructed. Such configurations exhibit inner cusp points, enabling accretion onto the central object, or outer cusp points, permitting outflows from the torus away from the central object. In contrast to neutral accretion disks, these charged configurations reveal additional closed equipotential surfaces above the equatorial plane: (i) along the polar axis, forming structures that take the place of the jets, which could be capable of scattering and polarizing light \citep{1993ARA&A..31..473A,Marin_2014}; and (ii) levitating off the equatorial plane, forming off-equatorial tori structures. These off-equatorial features may have significant astrophysical implications for the radiation emitted by accretion disks and the dynamics of oscillatory motion \citep{2008CQGra..25i5011K}. Notably, these unique configurations arise only when at least two of the following three effects are simultaneously present: an external magnetic field, a rotating central object, or a charged central object.

After establishing the background fields, let us discuss the assumption of a charged perfect fluid. First, to validate the use of our fluid model, we examine the upper limit for the applicability of kinetic theory, which typically applies to rarefied fluids with a number density of up to \(10^{24} \, \text{m}^{-3}\). In assuming particles of proton mass and a specific charge \(q_s \sim 10^{18}\), this implies a mass density limit of \(\rho_{\text{MHD}} \geq 10^{-3} \, \text{kg/m}^3\). However, it is important to emphasize that our fluid does not consist solely of proton plasma. The specific charge profiles within our disk are several orders of magnitude smaller, spanning six to nine orders lower. This average specific charge reflects a mixture of particles, such as a charged proton surrounded by a large number of neutral particles. Consequently, the conditions for the applicability of our model must be several orders higher than the \(\rho_{\text{MHD}}\)~limit. 

Secondly, one of the key assumptions in this model is the non-conductivity of the fluid, which represents the opposite limit of ideal magnetohydrodynamics (MHDs). While ideal MHD are a well established and highly effective approximation for many astrophysical systems, certain extreme scenarios, such as the merger of two neutron stars (NSs) or a neutron star--black hole (NS-BH) system, can produce plasmas with high temperatures and low densities. In such cases, non-ideal effects become significant. Recently, the scope of general relativistic MHDs (GRMHDs) has been expanded to include non-zero resistivity, referred to as non-ideal GRMHDs. This framework has now been  implemented in numerical simulation tools such as the Black Hole Accretion Code (BHAC) \citep{2019ApJS..244...10R}. Resistivity plays a critical role in magnetic reconnection and the growth of plasmoids in black hole accretion disk--jet systems. Through magnetic reconnection, the magnetic field topology changes, releasing magnetic energy that accelerates particles and produces non-thermal emissions, that are observed in various sources. Non-thermal emission remains a key uncertainty in ideal GRMHD models, particularly in Event Horizon Telescope (EHT) observations of accretion disks surrounding supermassive black holes. Studies such as \cite{2019ApJS..244...10R} have shown that high resistivity impacts system evolution by reducing turbulence induced by the magneto-rotational instability (MRI). These findings are consistent with earlier results from \cite{2017ApJ...834...29Q}. In this context, solutions assuming infinite resistivity like in this paper could provide valuable initial conditions for simulations exploring non-ideal effects.

The paper is organized as follows. In Section \ref{sec:level2}, we introduce and discuss the charged accretion disk model, including a discussion of the angular momentum distribution and the physical characteristics. In the following three sections, we present the results, first for the significant radii of the model, the innermost stable circular orbit (ISCO), and the cusp and center of the disk. We then proceed in Section \ref{sec:EqSol} with quantifying the shape and extension of the disk for a number of representative test cases, and in Section \ref{sec:5}, we proceed with the density distribution. % Please check that the intended meaning has been retained. %
 Finally, we summarize and discuss the results in Section \ref{sec:6}.

The equations are formulated in natural units, where both the speed of light, $c$, and the gravitational constant, $G$, are set to 1 ($c = G = 1$).
%%%%%%%%%%%%%%%%%%%%%%%%%%%%%%%%%%%%%%%%%%
\section{\label{sec:level2}The Charged Accretion Disk model}
\subsection{\label{sec:level21} General Equations}
A typical matter model requires specifying key properties of the fluid, such as its density, pressure, and velocity profile. Accretion disks are often treated as perfect fluids, assuming they behave as continuous media without any dissipative forces or viscosity, a simplification that offers useful insights into their macroscopic behaviour. In this study, we model the accretion disk as a charged perfect fluid, which we treat as a test fluid, implying that its self-gravity is negligible and does not alter the spacetime geometry. Such a setup was described before, e.g., in \cite{Kovar2016,Trova18}. In the following, we shortly collect the important equations for the convenience of the reader.

The basic equation governing the system is \cite{Trova18,misner1973gravitation}
\begin{align}
 \nabla_\beta \Tabmat=\Fabext J_\beta \, .
 \label{masterformular}
 \end{align}
 
    It is derived from the conservation of energy and Maxwell's equation using the decompositions $\Tab = \Tabmat + \Tabem$ for the energy--momentum tensor and $\Fab = \Fabext + \Fabint$ for the electromagnetic field tensor. Here, $\Fabext$ refers to electromagnetic fields external to the accretion disc, and $\Fabint$, to the internal field generated by the charged particle within the disk. We consider the internal field to be negligible, $\Fabint \ll \Fabext$. Although the external electromagnetic field is much stronger than the internal, it is still weak enough to leave the spacetime metric unaffected, as discussed in the Introduction. It is related to the electromagnetic vector potential $A_\nu$ as
\begin{equation}
\label{eq:ElectroMagTensor}
\Fabext = g^{\alpha\zeta} g^{\beta \nu}(\nabla_{\zeta}A_{\nu}-\nabla_{\nu}A_{\zeta})\,.
\end{equation}

    It is also assumed to be axially symmetric and stationary, leading to a specific form for the electromagnetic four-potential $A_{\zeta}=(A_t,0,0,A_\phi)$.

The four-current density vector, $J^{\alpha}$, depends on the charge density $\rho_q$, electrical conductivity $\sigma$, and four-velocity $U^{\alpha}$. Its form is described by Ohm’s law as 
\begin{equation}
 J^{\alpha}=\rho_qU^{\alpha}+\sigma \Fab U_{\beta}\,.
\end{equation}

We further assume that the matter exhibits circular motion around the black hole pulsar, defining the four-velocity component as $U^\alpha=(\utup,0,0,\upup)$. We then introduce the specific angular momentum  $\ell$ and  angular velocity $\Omega$, which can be expressed in terms of the four-velocity components as follows:
\begin{align}
\ell=-\frac{U_\phi}{U_t},\quad \quad \Omega=\frac{U^\phi}{U^t}\,, \quad \quad \Omega=-\frac{\ell\,g_{tt}+\gtp}{\ell\,\gtp+\gpp}\,.
\label{lomega}
\end{align}

\textcolor{black}{Usually, in ideal magnetohydrodynamics (MHDs), an infinite conductivity $\sigma$ is assumed, which necessitates that the electric field as seen by a comoving observer vanishes, $F^{\alpha\beta}U_\beta = 0$. However, this would lead to strong conditions on the electromagnetic vector potential \cite{Bonazzola1993}, which are, in general, not satisfied by our model. Therefore, we assume a finite conductivity. Moreover, the conditions  of axial symmetry and stationarity impose further constraints to ensure the integrability of the system. To satisfy our symmetry requirements, possible assumptions are to set $\sigma$ to zero or to assume a direction-dependent conductivity~\cite{Henriksen1974}, which will, however, both lead to the same final equations. We will assume $\sigma=0$, which leads to charged particles being tightly coupled to the fluid's motion, as assumed previously, e.g., in refs.~\cite{Kovar11,KovarTr14,Trova18}}

Based on these assumptions, the set of partial differential equations can be reformulated as follows \cite{Jaroszynski1980}:
\begin{subequations}
    \begin{align}
        &\partial_r W = \frac{\partial_r \gttup-2\ell \partial_r \gtpup+\ell^2\partial_r \gppup +q(\gttup-\ell \gtpup) \left(\partial_r \at+\Omega \partial_r \aphi \right)}{2(\gttup-\ell \gtpup)(1-\Omega\ell)}\label{eq:PartialDiff2a}\\ 
        &\partial_\theta W = \frac{\partial_\theta \gttup-2\ell \partial_\theta \gtpup+\ell^2\partial_\theta \gppup +q(\gttup-\ell \gtpup) \left(\partial_\theta \at+\Omega \partial_\theta \aphi \right)}{2(\gttup-\ell \gtpup)(1-\Omega\ell)},
        \label{eq:PartialDiff2b}
    \end{align}
    \label{eq:PartialDiff2ab}
\end{subequations}
with 
\begin{equation}
    \partial_{\alpha} W=\frac{\partial_{\alpha} p}{p+\epsilon}, \quad \text{with} \quad \alpha\in \left\{{r,t}\right\},
    \label{eq:WtoRho}
\end{equation}
where $W$ represents the effective potential. Here, 
$\gttup, \gtpup$, and $\gppup$ are the components of any general stationary axially symmetric spacetime. Nevertheless, in this study, the gravitational field is assumed to be generated by a Kerr black hole, described by the Kerr geometry in standard Boyer--Lindquist coordinates as follows:%\vspace{-12pt}
%\begin{adjustwidth}{-\extralength}{0cm}
%\centering %% If there is a figure in wide page, please release command \centering
\begin{align}
ds^2 =& -\left(1-\frac{2r}{\Sigma}\right) dt^2+\frac{\Sigma}{\Delta}dr^2+\Sigma d\theta^2+\left(r^2+a^2+\frac{2 r a^2}{\Sigma}\sin^2\theta\right)\sin^2\theta d\phi^2\nonumber\\
&-\frac{4r a \sin^2\theta}{\Sigma}dt d\phi,
\end{align}
%\end{adjustwidth}
where $\Sigma=r^2+a^2\cos^2\theta$, $\Delta=r^2-2r+a^2$, and $a$ represents the spin parameter of the black~hole. 

The term $q$ links the charge density of the disk to the pressure and energy density as~follows:
\begin{equation}
    q=2 \frac{\rho_q}{(p+\epsilon)\ut}
\end{equation}

    By  dividing Equations~(\ref{eq:PartialDiff2ab}a) and (\ref{eq:PartialDiff2ab}b), we obtain
\begin{equation}
    \frac{\mathrm{d}\theta}{\mathrm{d}r}=F(r,\theta),
\end{equation}
where $F(r,\theta)$ is determined in closed form under the following assumptions: (i) the gravitational field is predefined, (ii) the angular momentum distribution is specified, (iii) $q$ is arbitrarily chosen. In the absence of a magnetic field, the final term in each partial derivative disappears, reducing the system to a description consistent with the Polish doughnut~model.

\subsubsection{Angular Momentum Distribution}
\label{sec:AngMom}

In this study, we relax the assumption of constant angular momentum and instead adopt a distribution that combines a power-law dependence with a trigonometric form:
\begin{align}
\label{eq:AngMom}
 \ell(r,\theta)=
   \left\{
  \begin{array}{@{}ll@{}}
  \ell_{\mathrm{in}}\left(\sin{\theta}\right)^{2\delta}, & r<r_{ms,q}\\
  \ell_{\mathrm{in}}\left(\frac{L(r)}{\ell_{ms,q}}\right)^{\beta}\left(\sin{\theta}\right)^{2\delta}, & r\geq r_{ms,q}, 
    \end{array}\right.
\end{align}
with 
\begin{align}
   0\leq \beta\leq 1, \quad -1\leq \delta \leq 1.
\end{align}

The angular momentum at the inner edge $r_{\mathrm{in}}$ of the disk is denoted as \( \ell(r_{\mathrm{in}},\pi/2) = \ell_{\mathrm{in}} \). The choice of the inner radius of the disk is restricted to a range where bounded solutions with a cusp are possible. For an uncharged case, this corresponds to angular momentum values in the range \( \ell_{\rm ms} > \ell > \ell_{\rm mb} \), where \( \ell_{\rm ms} \) represents the marginally stable orbit, defined by the conditions \( \partial_r W_{q=0} = 0 \) and \( \partial_r^2 W_{q=0} = 0 \), and \( \ell_{\rm mb} \) corresponds to the marginally bounded orbit, where \( W_{q=0} = 0 \) and \( \partial_r W_{q=0} = 0 \). 

In the case of a charged fluid, since the explicit form of \( W \) is analytically unavailable, the inner edge of the disk is approximated to lie near the radius  \( r_{ms,q} \)  associated with the charged version of the marginally stable orbit, with angular momentum denoted as \( \ell_{ms,q} \). This value is determined by the conditions \( \partial_r W = 0 \) and \( \partial_r^2 W = 0 \). We defined a charged version of the ``Keplerian'' angular momentum \( L(r)\) in Equation~\eqref{eq:AngMom}, which is defined such that \( \partial_r W = 0 \). In the neutral case, the Keplerian angular momentum is determined as well by $\partial_r W_{q=0} = 0$. The inner radius is not entirely chosen arbitrarily. In the neutral case, it corresponds to an extremum of the effective potential \( W \), which means that $\ell(r_{\mathrm{in}},\pi/2)=L(r_{in})$, with $r_{in}<r_{ms,q}$. The chosen inner edge coincides with the cusp, which represents the location where material begins accreting onto the black hole. The center $r_c$ is determined using the same condition: $\ell_{c}=L(r_c)$ with $r_{c}>r_{ms,q}$.
 
The adopted angular momentum distribution, given by the Equation~\eqref{eq:AngMom}, is a slightly modified version of the model proposed in \cite{Qian2009}. In both distributions, the inner region of the disk exhibits constant angular momentum, while the central region transitions to super-Keplerian motion toward the center. Beyond this, the outer regions of the disk display sub-Keplerian motion. The authors of \cite{Qian2009} pointed out that the trigonometric distribution, which smoothly transitions between a constant angular momentum regime and a Keplerian profile, is a physically reasonable approximation. Numerical simulations support this assumption by showing that, at large radii, the specific angular momentum deviates slightly from the Keplerian value \cite{Fragile_2007,Fragile_2009,Sadowski_2008,Machida_2008}. In the inner disk region, spanning from the cusp (where accretion begins) to the ISCO, there is no universal agreement on the angular momentum profile. However, studies on axisymmetric accretion flows \cite{Proga_2003,Proga_2003b} suggest that this region often forms a pressure-supported torus in the equatorial plane, characterized by nearly constant specific angular momentum.

Regarding the role of the parameters, $\beta$ governs the radial behaviour of the angular momentum distribution, while $\delta$ influences its angular dependence. Since the inner edge and center of the disk are determined through the radial derivative in the equatorial plane, all quantities between these two critical radii, as well as those computed at these radii, are affected only by $\beta$. Consequently, $\delta$ does not contribute in this context. Furthermore, it is important to note that setting $\beta = \delta = 0$ recovers the case of a constant angular momentum~distribution.

\subsubsection{Magnetic Field and Charge Density}
As mentioned in the introduction, it has been shown that a black hole pulsar may create its own dipolar magnetic field. We argue that this magnetic field is a test field and it is not strong enough to perturb the metric. The electromagnetic potential for a dipole
magnetic test field in Boyer--Lindquist coordinates is given
by the following \cite{1975PhRvD..12.2218P,Prasanna1978}:
\begin{align}
&\At= -\frac{3}{2} \frac{a {\cal{M}}}{\xi^{2} \Sigma}\left(-\left(r-\cos ^{2}(\theta)\right)+\frac{1}{2 \xi} \ln \frac{r-1+\xi}{r-1-\xi}\times\left(r(r-1)+\left(a^{2}-r\right) \cos ^{2}(\theta)\right)\right) \\
&\Ap =-\frac{3}{4} \frac{{\cal{M}} \sin ^{2} \theta}{\xi^{2} \Sigma}\left((r-1) \Sigma+2 r\left(r+a^{2}\right)-\frac{1}{2 \xi} \ln \frac{r-1+\xi}{r-1-\xi}\left(\chi-4 r a^{2}\right)\right), 
\end{align}
where $ \xi=\sqrt{1-a^{2}}$ and $\chi(r, \theta)=\left(r^{2}+a^{2}\right)^{2}-\Delta a^{2} \sin ^{2} \theta$.
Here, ${\cal{M}}$ is the dipole moment of the external magnetic field, and we consider that ${\cal{M}}>0$ is a  $B-$field oriented along the symmetry axis $z$ positive. The frame-dragging effect in Kerr spacetime connects $\phi$- and $t$-components via cross terms in the metric. This leads to an $\at$-component in the description of the dipole magnetic field. This term will locally give rise to an electric part in the field.

The specific charge density of the fluid, along with the overall charge density, is treated as a free parameter in our model. The primary constraint imposed by our assumptions is that the magnetic field generated by the charged accretion disk must be significantly weaker, by at least an order of magnitude, than the external magnetic field. For simplicity and as a starting point for the analysis, we assume that the function $q=-k$ remains constant throughout the system. Under this assumption, the relationship between the pressure and the charge density is given by

\begin{equation}
    \rho_q=k (p+\epsilon)\vert U_t\vert.
    \label{eq:ChargeP}
\end{equation}

 The constant $k$ quantifies the overall degree of charge within the accretion disk. By assuming $q = -k$, a positively charged disk corresponds to $k > 0$, while a negatively charged disk corresponds to $k < 0$.
After normalizing the equations presented in \mbox{Equations (\ref{eq:PartialDiff2ab}a) and (\ref{eq:PartialDiff2ab}b)}, we introduce a new constant, $\mu=k\mathcal{M}$, that connects the disk's charge density with the dipole moment of the external dipole magnetic field. Then, $\mu < 0$ can mean either a positively charged fluid, embedded in a dipole magnetic field with a negative magnetic dipole, pointing along the $-z$-axis, or vice-versa. $\mu>0$ refers to a positively charged disk combined with a dipole moment pointing in the $+z-$axis or~vice-versa.

\subsection{Mass Density, Pressure and Energy Density}
Based on the chosen angular momentum distribution, the equipotential surfaces of the disk can be determined. These surfaces span the region between the inner edge and the center of the disk. Using the expression for the equipotential in the equatorial plane, derived from Equation (\ref{eq:PartialDiff2ab}a) and integrated between the inner edge (denoted by the subscript $\mathrm{in}$) and the center (denoted by the subscript $\mathrm{c}$), we have
\begin{equation}
    W_{\mathrm{Eq}}(r) = \int_{r_\mathrm{in}}^{r_\mathrm{c}} \partial_r W \, dr,
\end{equation}
which allows us to construct all equipotential surfaces in the disk, following the approach of \cite{SolerFont18}. From these equipotential surfaces, various disk properties can be derived by solving Equation \eqref{eq:WtoRho}.  

In this study, we consider adiabatic disks without radiation. For such disks, the energy density is related to the pressure and mass density by
\begin{equation}
    \epsilon = \rho + n p,
    \label{eq:EnergyP}
\end{equation}
where $n$ is the adiabatic constant. Furthermore, the fluid obeys an adiabatic pressure--mass density relationship:
\begin{equation}
    p = \kappa \rho^{\left(1 + \frac{1}{n}\right)}.
    \label{eq:PressureRho}
\end{equation}

By analysing the preceding equations in conjunction with Equation~\eqref{eq:WtoRho}, we can derive the expressions for the mass density, pressure, and energy density within the disk. By combining Equations~\eqref{eq:WtoRho}, \eqref{eq:EnergyP}, and \eqref{eq:PressureRho}, we obtain the following relationship between the effective potential $W$ and the rest mass density:
\begin{equation}
    \rho = \left(\frac{e^{W_{\rm in} - W} - 1}{\kappa(n+1)}\right)^n,
    \label{eq:RhoW}
\end{equation}
where $W_{\rm in}$ represents the value of the effective potential at the inner edge of the disk. The current analysis and Equations \eqref{eq:EnergyP} and \eqref{eq:PressureRho} provide explicit expressions for both the pressure and energy density. Using Equation \eqref{eq:ChargeP}, we further derive the profiles for the charge density. Importantly, the equipotential surfaces coincide with the surfaces of constant rest mass density, pressure, energy density, and charge density. Notably, the maximum values of all physical quantities occur precisely at the center of the disk.

\subsection{Interplay of $\kappa$ and $B$}
\label{sec:KappaB}
Through the selection of specific values for parameters $\mu$, $a$, $\beta$, and $\delta$ and choosing constant $k$, the equipotential surfaces can be constructed. Once these surfaces are defined, the rest mass density is obtained using Equation~\eqref{eq:RhoW} and the chosen value of $\kappa$. By choosing higher values of \(\kappa\), we obtain a more rarefied fluid. Subsequently, the pressure and the energy density are determined via Equations~\eqref{eq:PressureRho} and~\eqref{eq:EnergyP}, respectively. Finally, the charge density of the fluid is calculated using Equation~\eqref{eq:ChargeP} and the relation $k = \mu / B$. This process shows how $\kappa$ and $B$ together affect the magnitude of the fluid's charge and the induced magnetic field. We need to ensure that the choice of the $B$ value and \(\kappa\) value lead to an amplitude of the magnetic field produced by the disk itself that is not exceeding the external field, meaning that the assumption of negligible self-interaction is sustained. Consistent with our previous work and our background, we set \(\kappa = 10^4\) and $B \sim $\qty{e10}{\gauss}. These choices lead to solutions with a specific charge in the range of $ \rho_q/\rho \sim  10^6 - 10^8$, a mass density reaching $\rho \sim 10^{-19} - 10^{-21} $ (in SI units, $\rho= 1-100$ \qty{}{\kilogram\per\cubic\meter}), and a magnetic field amplitude induced by the disk attaining $B_d = 10 - 10^9$ \qty{}{\gauss}. Therefore, some solutions produce a magnetic field reaching $10^{-1}B$. These solutions are at the limit of our assumption of no~self-interaction.

\section{Innermost Stable Circular Orbit and Cusp and Center of the Disk}
\begin{figure}[H]
   % \centering
   \hspace{-1em} \begin{tabular}{ccc}
    \includegraphics[width=0.31\linewidth]{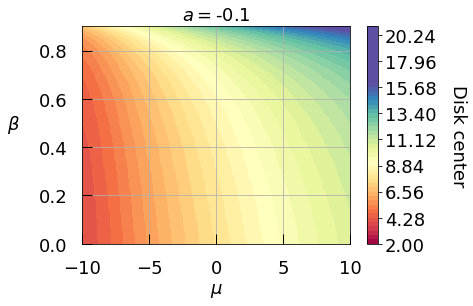}&
    \includegraphics[width=0.31\linewidth]{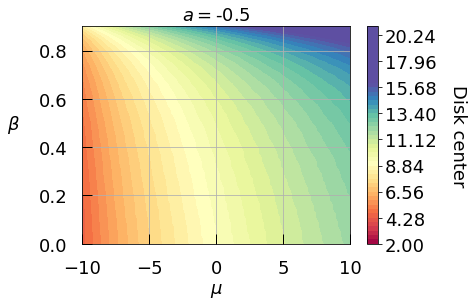} &
    \includegraphics[width=0.31\linewidth]{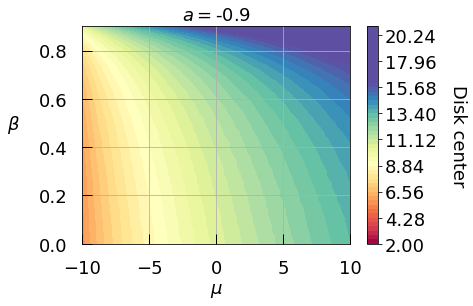} 
    \\
    \includegraphics[width=0.31\linewidth]{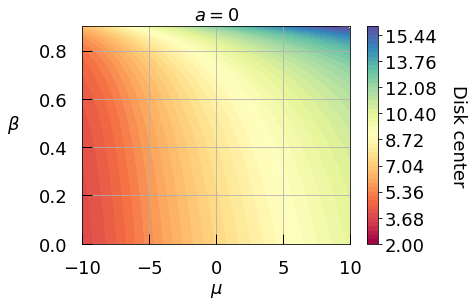} & \includegraphics[width=0.31\linewidth]{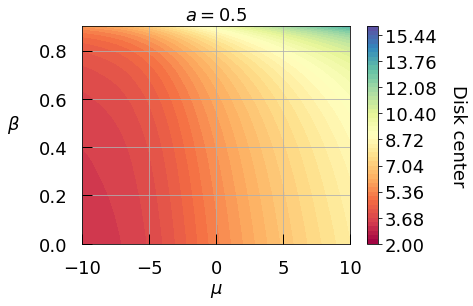} & \includegraphics[width=0.31\linewidth]{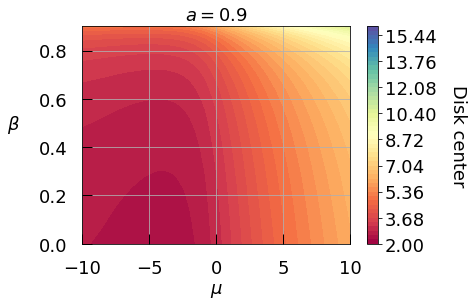} \\ 
    \end{tabular}\vspace{-9pt}
    \caption{Contour maps of the variation in the radius of the center of the disk with respect to the angular momentum parameter $\beta$ and the magnetic/charge parameter $\mu$ for prograde and retrograde spin.}
    \label{fig:Center}
\end{figure}

The first analysis is dedicated to the innermost stable circular orbit (ISCO), the cusp, and the center behaviour. Figure~\ref{fig:IscoInner} depicts the ISCO and the inner edge, and Figure~\ref{fig:Center} shows the behaviour of the disk center. 
 
\subsection{Innermost Stable Circular Orbit}
\begin{figure}[H]
   % \centering
  \hspace{-1em}  \begin{tabular}{cc}
      \includegraphics[width=0.45\hsize]{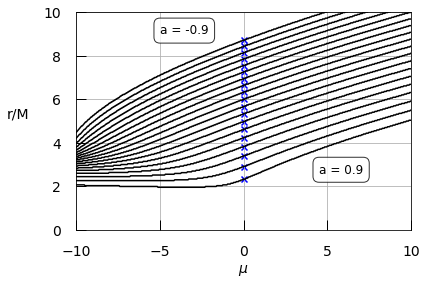}   &  \includegraphics[width=0.45\hsize]{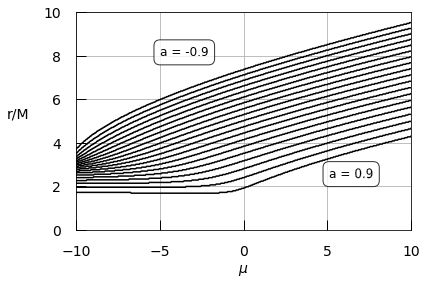}
    \end{tabular}\vspace{-9pt}
    \caption{Variation in the ISCO (\textbf{left panel}) and the disk's inner edge (\textbf{right panel}) with respect to the magnetic/charge parameter $\mu$ for various values of spin parameter $a$. The blue crosses show the values of the ISCO in the neutral case ($\mu = 0$).}
    \label{fig:IscoInner}
    \end{figure}

The left panel of Figure~\ref{fig:IscoInner} illustrates the variation in the normalized innermost stable circular orbit (ISCO) radius, $r/M$, with the magnetic/charge parameter, $\mu$, for black holes with different spin parameters, $a$, ranging from $a = 0$ (non-rotating Schwarzschild black hole) to $|a| = 0.9$ (rapidly rotating Kerr black hole).

    At $\mu = 0$, the ISCO radius corresponds to the gravitationally determined orbits, with $r/M = 6$ for $a=0$ and decreasing values for increasing spin parameters. This behaviour is consistent with the predictions of Kerr spacetime, where frame dragging reduces the ISCO radius for prograde orbits as $a$ increases and increases for retrograde orbits as $a$ decreases to negative values. The blue markers at $\mu = 0$ provide reference points for neutral particles, serving as benchmarks for gravitational ISCO radii in the absence of magnetic or charge~effects.
    \vspace{-6pt}

\vspace{-12pt}

    For $\mu \neq 0$, the ISCO radius is modified by the interaction between the particle’s charge or magnetic moment and the black hole’s electromagnetic field. The following behaviours are observed:

    \begin{enumerate}
    \item Schwarzschild Case ($a = 0$): The ISCO radius decreases monotonically with decreasing $\mu$, showing no qualitative change in behaviour.
    \item Kerr Case: While $r/M$ decreases with decreasing $\mu$ for small to moderate negative values of $\mu$, a reversal occurs at high negative $\mu$, where $r/M$ begins to increase. This change appears for highly rotating black holes with $a>0$. This non-monotonic behaviour suggests that for large negative $\mu$, the electromagnetic coupling introduces competing effects that counteract the inward pull on the ISCO radius. Moreover tapering behaviour for large negative $\mu$ appears and might indicate that the magnetic or charge interaction reaches a limiting effect, where further changes in $\mu$ no longer significantly affect the ISCO.
    \end{enumerate}

\subsection{Inner Edge}
The behaviour of the inner edge of the accretion disk (right panel of Figure~\ref{fig:IscoInner}) exhibits a striking resemblance to that of the innermost stable circular orbit (ISCO), with both regions sharing similar dynamic characteristics. Notably, the inner edge follows a comparable monotonic trend to that observed for the ISCO, reflecting a similar dependence on the physical parameters. The tapering effect is also present, which leads to the reduction in the distance between the ISCO and the inner edge.

\subsection{Center of the Disk} 

Figure \ref{fig:Center} illustrates the behaviour of the disk’s center. Regarding the effect of the black hole spin, for $a \leq 0$ and $a = 0.5$, the position of the disk center increases monotonically with both $\beta$ and $\mu$. This indicates that the center of the disk moves closer to the central mass as $\mu$ and $\beta$ decrease.

For a rapidly spinning black hole ($a = 0.9$), the behaviour is more complex. While the position of the disk center still exhibits a monotonic dependence on $\beta$, its variation with $\mu$ becomes non-monotonic. In particular, for highly negative values of $\mu$, the center is pushed farther away from the black hole. 

It is also observed that the disk center is more significantly influenced by variations in $\mu$ than by $\beta$, as the range of its displacement is greater when $\mu$ changes compared to $\beta$. This sensitivity to $\mu$ becomes less pronounced as the rotation rate of the black hole increases.

\section{Equipotential Surfaces, Aspect Ratio, and Radial Extension of the Disk}
\label{sec:EqSol}

\begin{figure}[H]
  %  \centering
    \includegraphics[width=0.3\linewidth]{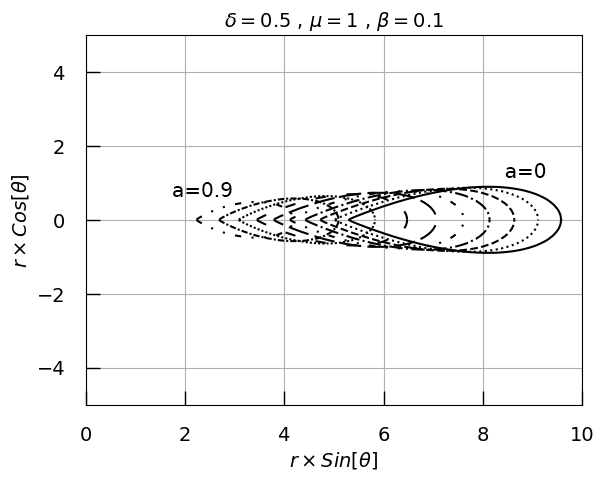}
    \includegraphics[width=0.32\linewidth]{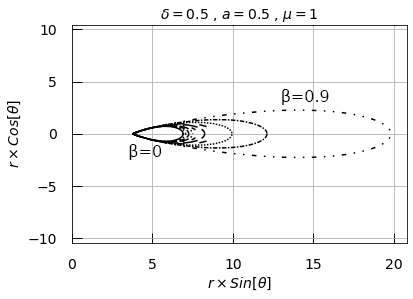}
    \includegraphics[width=0.32\linewidth]{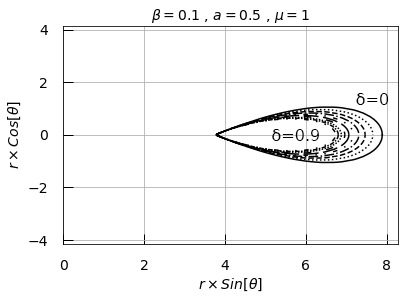}\vspace{-4pt}
    \caption{Shape of the charged accretion disk, defined as the zero-pressure surface for various combinations of the parameters. The (\textbf{leftmost}) panel shows the $C_{a}$ configuration. The (\textbf{middle})~panel presents the $C_{\beta}$ one. Finally, the (\textbf{rightmost}) panel shows the $C_{\delta}$ configuration.}
    \label{fig:SurfaceSpinBetaGamme}
\end{figure}

\begin{figure}[H]
 %   \centering
    \includegraphics[width=0.45\linewidth]{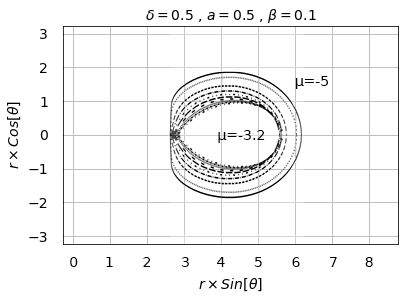}
    \includegraphics[width=0.45\linewidth]{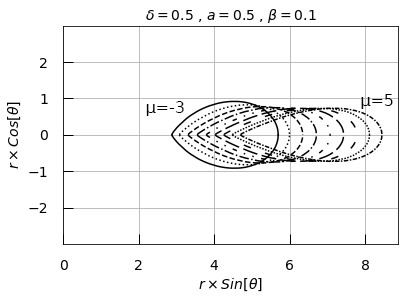}\vspace{-4pt}
    \caption{Same as Figure~\ref{fig:SurfaceSpinBetaGamme} but for configurations $C_{\mu,1}$ (\textbf{left}) and $C_{\mu,2}$ (\textbf{right}).}
    \label{fig:SurfaceMu}
\end{figure}

From this point onward, we will focus exclusively on positive or null values of $a$, as this scenario corresponds to the accretion disk extending closer to the black hole. This configuration is particularly relevant because it allows us to probe the extreme relativistic regime near the compact object, where strong gravitational effects dominate. Studying this region is crucial for gaining deeper insight into observational features such as high-energy emissions, relativistic jet formations, and other phenomena occurring in the innermost regions of the accretion flow. Then, we mostly focus on three significant values of the spin parameter, $a=0$, $a=0.5$, and $a=0.9$. These representative values cover both (close to) the vanishing spin as well as the highly spinning case close to the maximal value, with $a=0.5$ in between to better visualize the evolution. This section explores the shape of the disk, as illustrated in Figures~\ref{fig:SurfaceSpinBetaGamme} and \ref{fig:SurfaceMu}, which present the disk's surface under various conditions. Given the four key parameters influencing the disk’s geometry---spin $a$, angular momentum parameters $\beta$ and $\delta$, and charge/magnetic parameter $\mu$---we vary one parameter at a time while keeping the others fixed for each plot. We analyze five different configurations, described as follows:
\begin{enumerate}
    \item $C_{a}$: $\delta = 0.5,\mu = 1, \beta = 0.1$, and $a$ is varying in the range $[0,0.9]$.
    \item $C_{\beta}$: $a=0.5, \mu = 1, \delta=0.5$, and $\beta$ is varying in the range $[0,0.9]$.
    \item $C_{\delta}$: $a=0.5, \mu = 1, \beta= 0.1$, and $\delta$ is varying in the range $[0,0.9]$.
    \item $C_{\mu,1}$: $\delta = 0.5, a = 0.5, \beta=0.1$, and $\mu$ varying in the range $[-5,-3.2]$.
    \item $C_{\mu,2}$: $\delta = 0.5, a = 0.5, \beta=0.1$, and $\mu$ varying in the range $[-3,5]$.
\end{enumerate}

Figure~\ref{fig:SurfaceSpinBetaGamme} displays the disk surface for the configuration $C_{a}$, $C_{\beta}$, and $C_{\delta}$ from the leftmost to the rightmost panel, respectively.
\vspace{-4pt}
\vspace{-12pt}

Meanwhile, Figure~\ref{fig:SurfaceMu} presents the disk surface for different values of $\mu$ within the range $(-5,5)$ ($C_{\mu,1}$ and  $C_{\mu,2}$ configurations). Notably, the leftmost plot in Figure~\ref{fig:SurfaceMu} focuses on the range of high negative values ($\mu \in [-5, -3.2]$), highlighting the unusual disk geometry in this region, which resembles an almost bubble-like shape. In the rightmost plot, $\mu \in [-3, 5]$.

%%%%%%%%%%%%%%%%%%%%%%%%%%%%%%%%%%%%%%%%%%

\begin{figure}[H]
 %   \centering
    \includegraphics[width=0.45\linewidth]{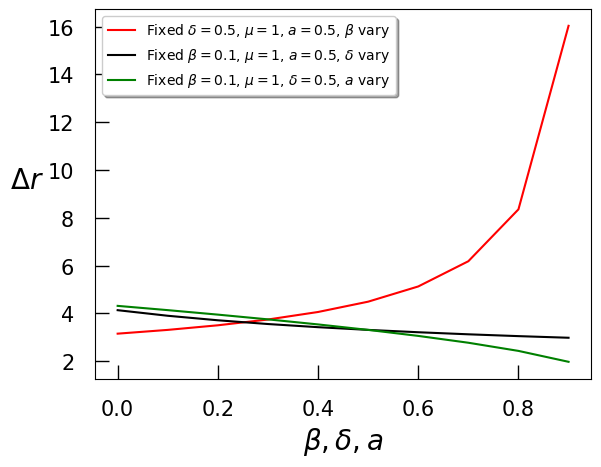}
    \includegraphics[width=0.45\linewidth]{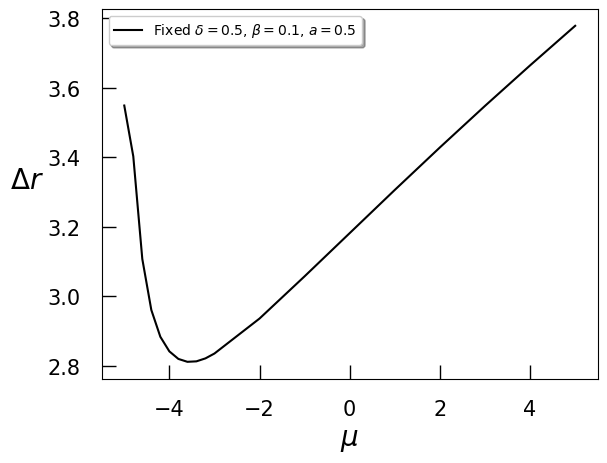}
    \caption{Variation in the $\Delta r$ ratio for the various parameters of the model. Configurations $C_{a}$, $C_{\beta}$, and $C_{\delta}$ are depicted on the (\textbf{left}).  The (\textbf{right}) panel presents configurations $C_{\mu,1}$ and $C_{\mu,2}$ together.}
    \label{fig:Radialextend}
\end{figure}

\begin{figure}[H]
  %  \centering
    \includegraphics[width=0.45\linewidth]{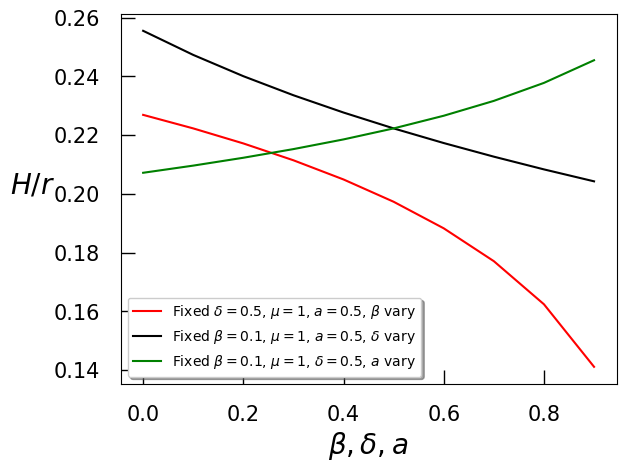}
    \includegraphics[width=0.45\linewidth]{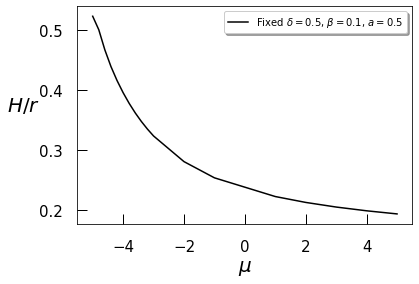}
    \caption{Same as Figure~\ref{fig:AspectRatio} but for the disk vertical scale height to radius.}
    \label{fig:AspectRatio}
\end{figure}

To facilitate the analysis of the disk's shape, we present two key quantities: the radial extent, $\Delta r = r_{\text{out}} - r_{\text{in}}$, and the disk vertical scale height to radius, $H/\Delta r$, which corresponds to the ratio of the distance from the equatorial plane to the highest $z-$altitude of the disk with the radial extension of the disk. These quantities are displayed in Figures~\ref{fig:Radialextend} and~\ref{fig:AspectRatio}, respectively. In each figure, the leftmost panel illustrates the variation in these quantities with respect to $a$, $\beta$, and $\delta$, as these parameters share a comparable range of variation (configurations $C_{a}$, $C_{\beta}$, and $C_{\delta}$). The rightmost panel highlights the variation exclusively with $\mu$ and both configurations $C_{\mu,1}$ and $C_{\mu,2}$ together, providing a focused view of its influence.
\vspace{-12pt}
\vspace{-12pt}

The disk shape observed in the present study is consistent with previous findings, displaying a typical structure across the entire range of positive values for $a$, $\beta$, $\delta$, and $\mu$, where the inner region of the disk is thinner than the outer part, making the disk section resemble an egg-shape section \cite{AbrJaSi78,Qian2009,SolerFont18}. Moreover, as summarized in \cite{Qian2009}, the solutions exhibit the typical characteristics of black hole accretion disks: (i) a funnel along the polar axis, relevant for studying jets; (ii) a pressure maximum, important for analyzing oscillatory motions; and (iii) a cusp point, corresponding to the self-crossing of two equipotential surfaces, which is relevant, for instance, for defining inner boundary conditions and where accretion can happen.
However, as $\mu$ becomes more negative, the disk shape undergoes a notable transformation resembling more of a bubble with a thicker inner region. Additionally, small values of $\beta$ lead to a smaller radial extent of the disk, while lower values of $a$ and $\delta$ result in a larger radial disk. This behaviour is emphasized in the leftmost plot of Figure~\ref{fig:Radialextend}. Conversely, the leftmost plot of Figure~\ref{fig:AspectRatio} shows that while smaller values of $a$ lead to a more radially extended disk, the disk becomes thinner compared to disks with higher values of $a$. In contrast, smaller values of $\beta$ result in a disk with a smaller radial extent but a thicker structure, which is opposite to the effect of increasing $\beta$. Similar trends are observed for $\delta$, where smaller values of $\delta$ produce disks that are both thicker and more radially extended. Regarding the variation with $\mu$, we observe that the radial extension of the disk exhibits a minimum around $\mu = -3.8$. In contrast, the aspect ratio behaves monotonically with respect to $\mu$. As $\mu$ increases, the disk transitions from being thick and less radially extended to being thin and more radially extended. This demonstrates a continuous evolution in the disk's structure as $\mu$ varies.

\section{Density Distribution and Amplitude}\label{sec:5}
In this section, we calculate the pressure, mass density, and charge density profiles derived from the solutions obtained in the previous sections. To perform these calculations, we define the parameters $n$, $\kappa$, and $\cal{M}$. For the initial setup, we choose a polytropic index of $n=3$ to model a relativistic fluid. Based on the discussion in the Introduction, we select the values $\kappa=\qty{e4}{}$ and $\cal{M}=\qty{4.24e-7}{}$, which yield a magnetic field magnitude of $B=\qty{4.23e-8}{}$ (in SI units, corresponding to $B=\qty{1.1e8}{\tesla}$) at $r=3M$~\citep{Kovar2016}.

\begin{figure}[H]
   % \centering
    \includegraphics[width=0.32\linewidth]{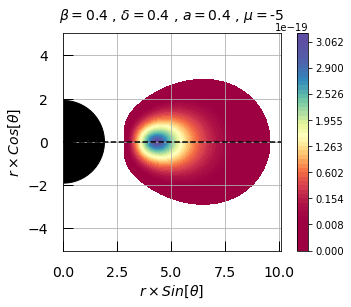}
    \includegraphics[width=0.32\linewidth]{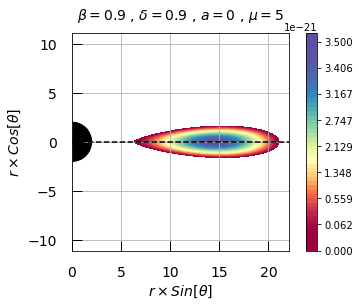}
    \includegraphics[width=0.32\linewidth]{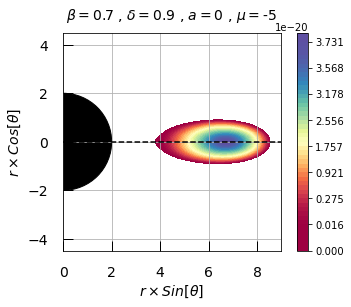}
    \caption{Density distribution in the entire space for some notable combination of the parameters.}
    \label{fig:densDist}
\end{figure}

Figure~\ref{fig:densDist} presents maps of the equi-mass density surfaces for selected disk solutions, highlighting those with notably distinct shapes.
\vspace{-6pt}

The solutions presented in Section~\ref{sec:EqSol} are characterized by their key physical properties, including the maximum rest mass density, $\rho_{\rm max}$, the maximum pressure, $p_{\rm max}$, and the induced magnetic field generated by the disk, $B_{\rm d}$. Following the approach in \cite{Kovar2016,Schroven2018}, to compare the strength of the ambient dominant magnetic field with the one generated by the charged rotating torus, we can estimate the order of its strength near the edge of the torus using the following simplification. In assuming a torus resembling an ideal mathematical regular torus, with a cross-section radius given by $(r_{\text{out}} - r_{\text{in}})/2$, and ``shrinking'' it figuratively into an infinitely thin charged ring at the torus center $r = r_c$, the magnetic field strength can be approximated as

\begin{equation}
   B_d \sim \frac{4\pi \times 10^{-7} \hat{I}}{2\pi (\hat{r}_{\text{out}} - \hat{r}_c)} \quad \text{(SI Unit)}, 
\end{equation}

Here, the constant $4\pi \times 10^{-7}$ represents the vacuum permeability in SI units, and $\hat{I} \sim \frac{\hat{Q} \hat{\omega}}{2\pi}$ is the estimated order of the total current passing through the cross-section of the original torus. In this equation the terms $\hat{X}$ denote quantities in SI Units.
The discussed physical quantities are summarized in Table~\ref{tab1} for the five configurations. They exhibit notable trends with respect to the system's parameters.

\begin{table}[H] 
\caption{Summary of disk solutions built in Section~\ref{sec:EqSol}.\label{tab1}}
%\newcolumntype{C}{>{\centering\arraybackslash}X}
\begin{tabularx}{\textwidth}{CCCCC}
\toprule
 &\boldmath{$C_{a}$}	& \boldmath{$C_{\beta}$}	& \boldmath{$C_{\delta}$} &\boldmath{$C_{\mu,1}$}+\boldmath{$C_{\mu,2}$} \\
\midrule
$\rho_{\rm max}$ (\qty{}{\kilogram\per\cubic\meter})   & 1.1--343 & 9--44 & 10 \textsuperscript{1}& 348--2  \\
$p_{\rm max}$ (\qty{e16}{\pascal}) & 0.01--25 & 0.2--1.6 & 0.2 \textsuperscript{1} & 26--0.27\\
$B_{\rm d}$ (\qty{}{\tesla}) & 0.7--46,400 & 8--4551 & 7--950& 218,700--750\\
\bottomrule
\end{tabularx}
\noindent{\footnotesize{\textsuperscript{1} No range for $C_{\delta}$, as $\delta$ does not affect the quantities in the equatorial plane from the inner edge to the center.}}
\end{table}

Specifically, $\rho_{\rm max}$, $p_{\rm max}$, and $B_{\rm d}$ all increase as parameters $a$, $\beta$, and $\delta$ are raised. Conversely, as $\mu$ transitions from highly negative to highly positive values, all three quantities consistently decrease. This behaviour underscores the significant influence of $\mu$, which encapsulates the electromagnetic coupling within the system, on the thermodynamic and magnetic properties of the disk. It is important to emphasize that $\rho_{\rm max}$ and $p_{\rm max}$ remain well within the range of validity for our fluid assumption, ensuring the physical consistency of the model. Furthermore, the disk-induced magnetic field, $B_{\rm d}$, is found to be orders of magnitude weaker than the external field, validating the approximation that the external magnetic field dominates the electromagnetic environment.

\begin{figure}[H]
   % \centering
        \includegraphics[width=0.32\linewidth]{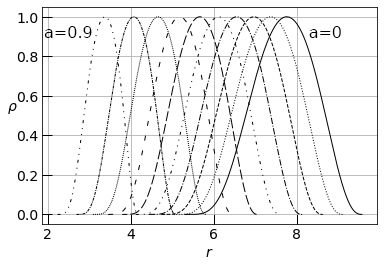} 
         \includegraphics[width=0.32\linewidth]{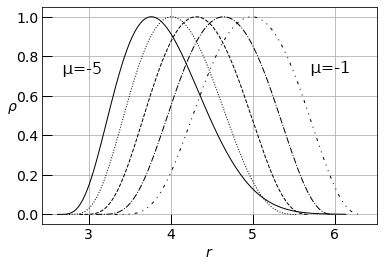}
      \includegraphics[width=0.32\linewidth]{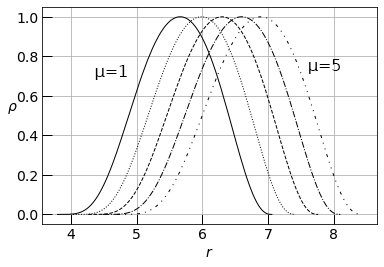}     
    \caption{Variation in the density distribution with $r$ for configuration $C_{a}$ (\textbf{leftmost plot}), for negative $\mu$ in the (\textbf{middle plot}), and for positive $\mu$ in the (\textbf{rightmost plot}).}
    \label{fig:RhoABetaGamma}
\end{figure}

\begin{figure}[H]
%    \centering
        \includegraphics[width=0.32\linewidth]{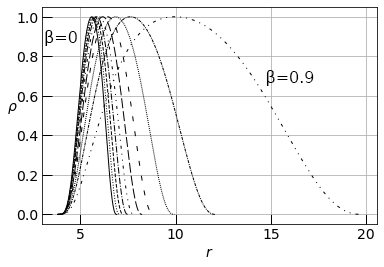} 
         \includegraphics[width=0.32\linewidth]{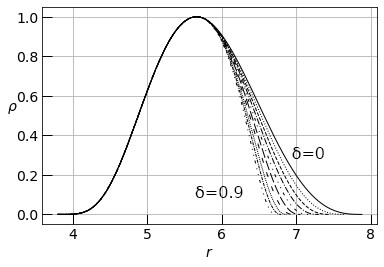}  
    \caption{Variation in the density distribution, normalized to $\rho_{max}$, with respect to $r$ for configurations $C_{\beta}$ (\textbf{left}) and $C_{\delta}$ (\textbf{right}).}
    \label{fig:RhoM}
\end{figure}

The last analysis is dedicated to the density distribution, which helps to understand how the matter is distributed inside the disk. Figure~\ref{fig:RhoABetaGamma} depicts $C_{a}$ and negative and positive $\mu$, from left to right, and Figure~\ref{fig:RhoM} presents $C_{\beta}$ and $C_{\delta}$, showing the variation in the rest mass density distribution normalized by its maximum value with all the parameters of the system. % Please check that the intended meaning has been retained. %
\vspace{-6pt}

We observe that the spin $a$ does not have a significant impact on the overall distribution of matter, apart from an overall shift in radius, as illustrated in the leftmost  panel of Figure~\ref{fig:RhoABetaGamma}. The middle and rightmost panels reveal that strongly negative $\mu$ tends to produce a distribution more concentrated in the inner regions of the disk, whereas positive $\mu$ shifts the distribution outward, similar to decreasing spin $a$.

Finally, Figure~\ref{fig:RhoM} reveals a noticeable influence of $\beta$ and $\delta$, the parameters that control the angular momentum distribution. 

For low values of $\beta$, the mass density falls off very steeply from the center of the disk and is radially almost symmetrical around the center. For higher values of $\beta$, the disk is more extended, and the mass density falls off quickly toward the inner edge, but trails off in a lower slope toward the outer edge.

Regarding $\delta$, this parameter primarily affects the outer regions of the disk. For low $\delta$ values, the density peak is more centrally located, and the matter is more evenly distributed across the disk. As $\delta$ increases, the outer edge of the disk shifts inward, resulting in a more compact structure. Overall, $\beta$ has a stronger influence on the density distribution than $\delta$. Figure~\ref{fig:densDist} highlights three representative solutions: a disk with the density peak near the inner edge with a highly concentration of matter around the center, a disk with a density peak in the middle of the disk less concentrated in the center, and a disk with the density peak near the outer edge. When $\beta$ and $\delta$ are equal, the effect of $\beta$ dominates. Only when $\beta$ decreases does the influence of $\delta$ become apparent.

%%%%%%%%%%%%%%%%%%%%%%%%%%%%%%%%%%%%%%%%%%
\section{Discussion}\label{sec:6}

In this section, we discuss the influence of each of the four parameters and their contributions separately: the spin $a$, the parameter $\mu$ (related to the charge and magnetic field), and the angular momentum parameters, $\beta$ (radial dependence) and $\delta$ (angular~dependence).

\subsection{The spin $a$ of the Black Hole}
The influence of the spin on the equilibrium structures of relativistic charged or neutral tori is well documented, and our analysis aligns with previous studies. The inner and outer edges of the disk closely follow the behaviour of the ISCO. As the spin increases, the disk becomes more compact, with all three critical radii moving closer to the black hole. This results in smaller, thicker disks as the black hole rotates faster.

\subsection{The Charge/Magnetic Parameter $\mu$}
The parameter $\mu$, linking the dipole magnetic field of the black hole and the charge of the disk significantly influences the charged particles and the structure of the accretion disk. Comparing with the neutral fluid case ($\mu = 0$), we observe the following:
\begin{itemize}
    \item Negative $\mu$: Negatively charged disks, when combined with a positively aligned external magnetic field (or vice versa), become thicker and more extended if $\mu$ is highly negative. The disk is closer to the compact object, with a denser matter distribution concentrated near the inner edge and produces a stronger induced magnetic field.
    \item Positive $\mu$: Positively charged disks, when combined with a positively aligned external magnetic field (or both negative), are thinner, located further from the central source, and less extended if $\mu$ is near neutral. If $\mu$ is highly positive, the disks become more extended, with matter concentrated in its outer regions.
\end{itemize}

\subsection{Angular Momentum Parameters $\beta$ and $\delta$}
Parameters $\beta$ and $\delta$ represent, respectively, the radial and angular dependencies of the angular momentum distribution. For $\beta = \delta = 0$, the angular momentum is constant, recovering the classical case of a constant angular momentum disk. When $\beta$ is strong relative to $\delta$ or equivalent, the disks are less extended, closer to the compact object, and thinner, with a matter distribution concentrated near the inner edge than in the constant~case.

As $\beta$ approaches $1$, and $\delta=0$, the angular momentum distribution approximates a Keplerian profile. In this regime, the pressure gradient diminishes, becoming nearly constant in both $r$ and $\theta$. Consequently, the disk flattens and aligns with the equatorial plane, consistent with expectations for a Keplerian accretion disk.

\subsection{Combined Parameters Analysis}
Finally, analyzing the combined influence of the parameters reveals several key interdependencies. The radial dependence ($\beta$) dominates the angular dependence ($\delta$) in determining the matter distribution and the location of the disk's central concentration. High $\beta$, negative $\mu$, and large $a$ result in disks that are closer to the black hole, with a denser matter distribution near the inner edge. This configuration could enhance the accretion rate and the production of high-energy radiation near the event horizon, potentially leading to a more luminous system. \textcolor{black}{Also, due to the interdependencies of the parameters, a certain degeneracy is introduced. For example, it should be generally possible to mimic the shape and features of accretion disks around spinning black holes without an external field by choosing appropriate values for the dipole/magnetic coupling $\mu$ and the angular momentum distribution parameters. Therefore, neglecting the possibility of a black hole with a dipole magnetic field may have an impact on estimations of black hole spins from accretion disk observations.}

\subsection{Comparison with Results from \cite{Schroven2018}}
In \cite{Schroven2018}, the focus was on finding analytical solutions to the two partial differential equations (Equations~(\ref{eq:PartialDiff2ab}a) and (\ref{eq:PartialDiff2ab}b)). To achieve this, the fluid was assumed to exhibit a specific charge and rotate in rigid rotation, i.e., with a constant angular velocity $\Omega$. Under these conditions, the resulting disks with a rest mass density and induced magnetic field, within the acceptable physical range, exhibited radial and vertical extensions that were significantly smaller than those observed in the disks constructed in this work. However, solutions featuring an outer cusp could be obtained in their framework, which is not the case in our setup due to the chosen angular momentum distribution.

%\subsection*{}
In conclusion, we demonstrated that thick accretion disks in equilibrium in such environments can exist and reproduce the classical features observed in the standard neutral model of accretion disks. These include the presence of a pressure maximum, a funnel along the polar axis where jets can happen, and a cusp where accretion can occur. However, the overall shape of the disk, especially for highly negatively charged tori with a positively aligned external dipolar magnetic field, deviates from the standard egg-shaped configurations, leading to a more toroidal-like structure. Furthermore, a non-constant angular momentum distribution results  some combination of the parameters in disks where matter is more concentrated in the outer regions of the system. The next step will involve testing the stability of such solutions and refining our assumption of a constant ratio between the charge and rest mass densities to better reflect the physical reality of such~systems.
%%%%%%%%%%%%%%%%%%%%%%%%%%%%%%%%%%%%%%%%%%
% \section{Conclusions}
%\vspace{-6pt}
% This section is not mandatory, but can be added to the manuscript if the discussion is unusually long or complex.

%%%%%%%%%%%%%%%%%%%%%%%%%%%%%%%%%%%%%%%%%%
%\section{Patents}

%This section is not mandatory, but may be added if there are patents resulting from the work reported in this manuscript.

%%%%%%%%%%%%%%%%%%%%%%%%%%%%%%%%%%%%%%%%%%
\vspace{6pt} 

%%%%%%%%%%%%%%%%%%%%%%%%%%%%%%%%%%%%%%%%%%
%% optional
%\supplementary{The following supporting information can be downloaded at:  \linksupplementary{s1}, Figure S1: title; Table S1: title; Video S1: title.}

% Only for journal Methods and Protocols:
% If you wish to submit a video article, please do so with any other supplementary material.
% \supplementary{The following supporting information can be downloaded at: \linksupplementary{s1}, Figure S1: title; Table S1: title; Video S1: title. A supporting video article is available at doi: link.}

% Only for journal Hardware:
% If you wish to submit a video article, please do so with any other supplementary material.
% \supplementary{The following supporting information can be downloaded at: \linksupplementary{s1}, Figure S1: title; Table S1: title; Video S1: title.\vspace{6pt}\\
%\begin{tabularx}{\textwidth}{lll}
%\toprule
%\textbf{Name} & \textbf{Type} & \textbf{Description} \\
%\midrule
%S1 & Python script (.py) & Script of python source code used in XX \\
%S2 & Text (.txt) & Script of modelling code used to make Figure X \\
%S3 & Text (.txt) & Raw data from experiment X \\
%S4 & Video (.mp4) & Video demonstrating the hardware in use \\
%... & ... & ... \\
%\bottomrule
%\end{tabularx}
%}

%%%%%%%%%%%%%%%%%%%%%%%%%%%%%%%%%%%%%%%%%%
\authorcontributions{Conceptualization, methodology and writing---original draft preparation, Audrey Trova; Conceptualization, methodology, writing---review and editing, Eva hackmann; %MDPI: For research articles with several authors, a short paragraph specifying their individual contributions must be provided. The following statements should be used ``Conceptualization, X.X. and Y.Y.; methodology, X.X.; software, X.X.; validation, X.X., Y.Y. and Z.Z.; formal analysis, X.X.; investigation, X.X.; resources, X.X.; data curation, X.X.; writing---original draft preparation, X.X.; writing---review and editing, X.X.; visualization, X.X.; supervision, X.X.; project administration, X.X.; funding acquisition, Y.Y. All authors have read and agreed to the published version of the manuscript.'', please turn to the  \href{http://img.mdpi.org/data/contributor-role-instruction.pdf}{CRediT taxonomy} for the term explanation. Authorship must be limited to those who have contributed substantially to the work~reported.
}

\funding{This research was funded by the Deutsche Forschungsgemeinschaft (DFG, German Research Foundation) grant number 510727404 and by DFG in the context of Germany’s Excellence Strategy---EXC-2123 QuantumFrontiers grant number 390837967. %MDPI: Please add: ``This research received no external funding'' or ``This research was funded by NAME OF FUNDER grant number XXX.'' and  and ``The APC was funded by XXX''. Check carefully that the details given are accurate and use the standard spelling of funding agency names at \url{https://search.crossref.org/funding}, any errors may affect your future funding.
}

\dataavailability{Data sharing is not applicable (only appropriate if no new data is generated or the article describes entirely theoretical research) %MDPI: We encourage all authors of articles published in MDPI journals to share their research data. In this section, please provide details regarding where data supporting reported results can be found, including links to publicly archived datasets analyzed or generated during the study. Where no new data were created, or where data is unavailable due to privacy or ethical restrictions, a statement is still required. Suggested Data Availability Statements are available in section ``MDPI Research Data Policies'' at \url{https://www.mdpi.com/ethics}.
} 

% Only for journal Nursing Reports
%\publicinvolvement{Please describe how the public (patients, consumers, carers) were involved in the research. Consider reporting against the GRIPP2 (Guidance for Reporting Involvement of Patients and the Public) checklist. If the public were not involved in any aspect of the research add: ``No public involvement in any aspect of this research''.}

% Only for journal Nursing Reports
%\guidelinesstandards{Please add a statement indicating which reporting guideline was used when drafting the report. For example, ``This manuscript was drafted against the XXX (the full name of reporting guidelines and citation) for XXX (type of research) research''. A complete list of reporting guidelines can be accessed via the equator network: \url{https://www.equator-network.org/}.}

% Only for journal Nursing Reports
%\useofartificialintelligence{Please describe in detail any and all uses of artificial intelligence (AI) or AI-assisted tools used in the preparation of the manuscript. This may include, but is not limited to, language translation, language editing and grammar, or generating text. Alternatively, please state that “AI or AI-assisted tools were not used in drafting any aspect of this manuscript”.}

\acknowledgments{We acknowledge the Deutsche Forschungsgemeinschaft (DFG, German Research Foundation) for funding the research leading to this project.}%under project number 510727404. EH is grateful for support from the DFG under Germany’s Excellence Strategy---EXC-2123 QuantumFrontiers---390837967.}

\conflictsofinterest{The authors declare no conflicts of interest. %MDPI: Declare conflicts of interest or state ``The authors declare no conflicts of interest.'' Authors must identify and declare any personal circumstances or interest that may be perceived as inappropriately influencing the representation or interpretation of reported research results. Any role of the funders in the design of the study; in the collection, analyses or interpretation of data; in the writing of the manuscript; or in the decision to publish the results must be declared in this section. If there is no role, please state ``The funders had no role in the design of the study; in the collection, analyses, or interpretation of data; in the writing of the manuscript; or in the decision to publish the results''.
} 

\begin{adjustwidth}{-\extralength}{0cm}
%\printendnotes[custom] % Un-comment to print a list of endnotes
\printendnotes[custom]

\reftitle{References}

\PublishersNote{}
\end{adjustwidth}
\end{document}